\documentstyle[12pt]{article}

\begin{document}

\baselineskip=14pt plus 1pt minus 1pt

\begin{center}{\large \bf
Unified description of magic numbers of metal clusters in terms of 
the 3-dimensional $q$-deformed harmonic oscillator}
\bigskip\bigskip

{Dennis Bonatsos$^{\#}$,
N. Karoussos$^{\#}$,
D. Lenis$^{\#}$, 
P. P. Raychev$^\dagger$,
R. P. Roussev$^\dagger$
P. A. Terziev$^\dagger$
\bigskip

{$^{\#}$ Institute of Nuclear Physics, N.C.S.R.
``Demokritos''}

{GR-15310 Aghia Paraskevi, Attiki, Greece}

{$^\dagger$ Institute for Nuclear Research and Nuclear Energy, Bulgarian
Academy of Sciences }

{72 Tzarigrad Road, BG-1784 Sofia, Bulgaria}}

\end{center}

\bigskip\bigskip
\centerline{\bf Abstract} \medskip
Magic numbers predicted by a 3-dimensional $q$-deformed harmonic oscillator 
with u$_q$(3) $\supset$ so$_q$(3) symmetry are compared to experimental 
data for atomic clusters of alkali metals (Li, Na, K, Rb, Cs), noble metals
(Cu, Ag, Au), divalent metals (Zn, Cd), and trivalent metals (Al, In),  
as well as to theoretical predictions 
of jellium models, Woods--Saxon and wine bottle potentials, and to the 
classification scheme using the $3n+l$ pseudo quantum number. In alkali 
metal clusters and noble metal clusters the 
3-dimensional $q$-deformed harmonic oscillator correctly predicts all
experimentally observed magic numbers up to 1500 (which is the expected 
limit of validity for theories based on the filling of electronic shells), 
while in addition it gives satisfactory results for the magic numbers of 
clusters of divalent metals and trivalent metals, 
thus indicating that u$_q$(3), 
which is a nonlinear extension of the u(3) symmetry of the spherical 
(3-dimensional isotropic) harmonic oscillator, is a good candidate for 
being the symmetry of systems of several  metal clusters.  
The Taylor expansions of angular momentum dependent potentials approximately 
producing the same spectrum as the 3-dimensional $q$-deformed harmonic 
oscillator are found to be similar to the Taylor expansions of the 
symmetrized Woods--Saxon and ``wine-bottle'' symmetrized Woods--Saxon 
potentials, which are known to provide successful fits of the Ekardt 
potentials.


\newpage

{\bf 1. Introduction} 

Metal clusters have been recently the subject of many investigations
(see \cite{deHeer,Brack,Nester,deHeer2} for relevant reviews). 
One of the first 
fascinating findings in their study was the appearance of magic numbers, 
analogous to 
but different from the magic numbers appearing in the shell structure of 
atomic nuclei \cite{Mayer}. Different kinds of metallic clusters [alkali
metals (Na \cite{Martin,Bjorn,Knight1,Peder}, Li \cite{Brec,Brec2}, 
K \cite{Knight2},
Rb \cite{Bhaskar}, Cs \cite{Bjorn,Goehlich,Bergmann}), noble metals 
(Cu \cite{Kata1,Knick}, Ag \cite{Kata1,Alam}, Au \cite{Kata1}), 
divalent metals of the IIB group (Zn, Cd) \cite{Kata2}, 
trivalent metals of the III group (Al, In) \cite{Persson}] 
exhibit different sets of magic numbers. 
The analogy between the magic numbers observed in metal clusters and the 
magic numbers observed in atomic nuclei 
led to the early description of metal 
clusters in terms of the Nilsson--Clemenger model \cite{Clem},
which is a simplified version of the Nilsson model \cite{Nilsson1,Nilsson2} 
of atomic 
nuclei, in which no spin-orbit interaction is included. Further theoretical
investigations in terms of the jellium model \cite{Ekardt,Beck} 
demonstrated that the mean field potential in the case of simple metal 
clusters bears great similarities to the Woods--Saxon potential 
of atomic nuclei, with a slight modification of the ``wine bottle''
type \cite{Kotsos,Kotsos2}. 
The Woods--Saxon potential itself looks like a harmonic 
oscillator truncated at a certain energy value and flattened at the bottom. 
It should also be recalled that an early schematic explanation of the 
magic numbers of metallic clusters has been given in terms of a scheme 
intermediate between the level scheme of the 3-dimensional harmonic 
oscillator and the square well \cite{deHeer}. Again in this case the 
intermediate 
potential resembles a harmonic oscillator flattened at the bottom.  

On the other hand, modified versions of harmonic oscillators 
\cite{Bie,Mac} have been 
recently investigated in the novel mathematical framework of quantum algebras 
\cite{Chari,Klimyk}, which are nonlinear generalizations of the usual Lie
algebras. The spectra of $q$-deformed oscillators increase either 
less rapidly (for $q$ being a phase factor, i.e. $q=e^{i\tau}$ with 
$\tau$ being real)
or more rapidly (for $q$ being real, i.e. $q=e^{\tau}$ with $\tau$ being 
real)
in comparison to the equidistant spectrum 
of the usual harmonic oscillator \cite{Roman}, while the corresponding 
(equivalent within the limits of perturbation theory or WKB-equivalent) 
potentials \cite{BDKPRA,BDKJMP}
resemble the harmonic oscillator potential,
truncated at a certain energy (for $q$ being a phase factor) 
or not (for $q$ being real), 
the deformation inflicting an overall
widening or narrowing of the potential, depending on the value of the 
deformation parameter $q$.   

Very recently, a $q$-deformed version of the 3-dimensional harmonic 
oscillator has been constructed \cite{Terziev}, taking advantage of the 
u$_q$(3) $\supset$ so$_q$(3) symmetry \cite{Smirnov,Jeugt}. 
The spectrum of this 3-dimensional $q$-deformed harmonic oscillator 
has been found \cite{Terziev} to reproduce very well the spectrum of the 
modified harmonic oscillator introduced by Nilsson 
\cite{Nilsson1,Nilsson2}, without the 
spin-orbit interaction term. Since the Nilsson model without the 
spin orbit term is essentially the Nilsson--Clemenger model used 
for the description of metallic clusters \cite{Clem}, it is worth examining 
if the 3-dimensional $q$-deformed harmonic oscillator can reproduce 
the magic numbers of simple metallic clusters and, in the case that this is 
possible, to determine potentials giving the same spectrum as this 
oscillator and compare them to the symmetrized Woods--Saxon and 
``wine-bottle'' symmetrized Woods--Saxon potentials, which successfully fit 
\cite{Kotsos,Kotsos2} the Ekardt potentials \cite{Ekardt}. 
These are the subjects of the present investigation. 

It is worth mentioning at this point that an effort has been made to describe 
the magic numbers of metal clusters by a quantum number $3n+l$ \cite{Martin}, 
where $n$
is the number of nodes in the solution of the radial Schr\"odinger equation 
and $l$ is the angular momentum quantum number. This approach has been 
inspired by the fact that degenerate energy levels in the hydrogen atom 
are characterized by the same value of the quantum number $n+l$, due to the 
so(4) symmetry underlying this system, while 
degenerate energy levels in the spherical harmonic oscillator (i.e. the 
3-dimensional isotropic harmonic oscillator) are characterized by the same 
value of the parameter $2n+l$, due to the su(3) symmetry underlying 
this system.  The $3n+l$ quantum number has been used \cite{Martin}
to approximate the magic numbers of alkali metal clusters with some success, 
and focusing potentials characterized by this degeneracy have been 
determined \cite{Ostrovsky},  
but no relevant Lie symmetry could be determined \cite{Ostrovsky,Koch}.

In Section 2 the 3-dimensional $q$-deformed harmonic oscillator will
be briefly described, while in Section 3 the magic numbers provided 
by this oscillator will be compared with the experimental data for Na and Li
clusters, as well as with the predictions of other theories (jellium model, 
Woods--Saxon and ``wine bottle'' potentials, classification scheme using the 
$3n+l$ pseudo quantum number). Additional comparisons of magic numbers 
predicted by the 3-dimensional $q$-deformed harmonic oscillator to 
experimental data and to the results of other theoretical approaches 
will be made in Section 4 (for other alkali metal clusters 
 and noble metal clusters), 
Section 5 (for divalent group IIB metal clusters), and Section 6 (for 
trivalent group III metal clusters), while in Section 7 potentials 
giving approximately the same spectrum as the 3-dimensional $q$-deformed 
harmonic oscillator will be determined and subsequently compared to the 
symmetrized Woods--Saxon and ``wine-bottle'' symmetrized Woods--Saxon 
potentials. Finally, Section 8 will contain
discussion of the present results and plans for further work. 

{\bf 2. The 3-dimensional $q$-deformed harmonic oscillator} 

The space of the 3-dimensional $q$-deformed harmonic oscillator consists of 
the completely symmetric irreducible representations of the quantum algebra
u$_q$(3). In this space a deformed angular momentum algebra, so$_q$(3), 
can be defined \cite{Terziev}. 
The Hamiltonian of the 3-dimensional $q$-deformed 
harmonic oscillator is defined so that it satisfies the following 
requirements:

a) It is an so$_q$(3) scalar, i.e. the energy is simultaneously measurable
with the $q$-deformed  angular momentum related to the algebra so$_q$(3) 
and its $z$-projection.   

b) It conserves the number of bosons, in terms of which the quantum 
algebras u$_q$(3) and so$_q$(3) are realized. 

c) In the limit $q\to 1$ it is in agreement with the Hamiltonian of the usual 
3-dimensional harmonic oscillator. 
 
It has been proved \cite{Terziev} that the Hamiltonian of the 3-dimensional 
$q$-deformed harmonic oscillator satisfying the above requirements 
takes the form
\begin{equation}
H_q = \hbar \omega_0 \left\{ [N] q^{N+1} - {q(q-q^{-1})\over [2] } C_q^{(2)}
\right\},
\end{equation}
where $N$ is the number operator and $C_q^{(2)}$ is the second order 
Casimir operator of the algebra so$_q$(3), while 
\begin{equation}
[x]= {q^x-q^{-x} \over q-q^{-1}}
\end{equation} 
is the definition of $q$-numbers and $q$-operators. 

The energy eigenvalues of the 3-dimensional $q$-deformed harmonic oscillator 
are then \cite{Terziev}
\begin{equation}
E_q(n,l)= \hbar \omega_0 \left\{ [n] q^{n+1} - {q(q-q^{-1}) \over [2]}
[l] [l+1] \right\}, 
\end{equation}
where $n$ is the number of vibrational quanta and $l$ is the eigenvalue of
the 
angular momentum, obtaining the values
$l=n, n-2, \ldots, 0$ or 1.  

In the limit of $q\to 1$ one obtains ${\rm lim}_{q\to 1} E_q(n,l)=
\hbar \omega_0 n$, which coincides with the classical result. 

For small values of the deformation parameter $\tau$ (where $q=e^{\tau}$)
one can expand eq. (3) in powers of $\tau$  obtaining \cite{Terziev}
$$
E_q(n,l)= \hbar \omega_0 n -\hbar \omega_0 \tau \left(l(l+1)-n(n+1)\right)
$$
\begin{equation}
-\hbar \omega_0 \tau^2 \left( l(l+1)-{1\over 3} n(n+1)(2n+1) \right)
+ {\cal O} (\tau^3).
\end{equation}

The last expression to leading order bears great similarity to the modified 
harmonic 
oscillator suggested by Nilsson \cite{Nilsson1,Nilsson2} 
(with the spin-orbit term omitted)
\begin{equation}
V= {1 \over 2} \hbar \omega \rho^2 -\hbar \omega \kappa' 
({\bf L}^2 - <{\bf L}^2>_N ), \qquad \rho=r \sqrt {M\omega \over \hbar} ,
\end{equation}
where 
\begin{equation}
<{\bf L}^2>_N = {N(N+3)\over 2}.
\end{equation}
The energy eigenvalues of Nilsson's  modified harmonic oscillator
are \cite{Nilsson1,Nilsson2}
\begin{equation}
E_{nl}= \hbar \omega n -\hbar \omega \mu' \left( l(l+1)-{1\over 2}
n(n+3)\right). 
\end{equation}
It has been proved \cite{Terziev} that the spectrum of the 3-dimensional 
$q$-deformed harmonic oscillator closely reproduces the spectrum of 
the modified harmonic oscillator of Nilsson. In both cases the effect 
of the $l(l+1)$ term is to flatten the bottom of the harmonic 
oscillator potential, thus making it to resemble the Woods--Saxon 
potential. 

The level scheme of the 3-dimensional $q$-deformed harmonic oscillator 
(for $\hbar \omega_0 =1$ and $\tau = 0.038$) is given in Table 1, up to 
a certain energy. Each level is characterized by the quantum numbers 
$n$ (number of vibrational quanta) and $l$ (angular momentum). Next 
to each level its energy, the number of particles it can accommodate
(which is equal to $2(2l+1)$) and the total number of particles up to 
and including this level are given. If the energy difference between 
two successive levels, which we shall denote by $\delta$,
is larger than 0.39, it is  considered as a gap 
separating two successive shells and the energy difference is reported 
between the two levels. In this way magic numbers can be easily read 
in the table: they are the numbers appearing above the gaps, written in 
boldface characters. 

Additional level schemes of the 3-dimensional $q$-deformed harmonic
oscillator 
are given in Table 2 (for $\tau=0.020$ and energy gap $\delta =0.20$) and in 
Table 3 (for $\tau=0.050$ and energy gap $\delta =0.38$). 
The following remarks are now in place:

i) Small magic numbers do not change much as the parameter $\tau$ is varied
(taking positive values), while large magic 
numbers get more influenced by the parameter modification. 
In general, the ordering of the levels does not change rapidly with the 
value of the parameter $\tau$ (for $\tau>0$). 

ii)  Rapid change of the magic numbers as a function of $\tau$ occurs 
when $\tau$ takes negative values, but this case is irrelevant to the 
contents of the present work. 

iii) Magic numbers are influenced nore drastically by the value of the 
energy gap $\delta$. If in the spectrum obtained for a given value 
of the parameter $\tau$ the energy separation between two successive 
levels is only slightly smaller than the energy gap $\delta$, 
this can be considered as an indication of the presence of a ``secondary''
magic number. (See the end of Section 3 for specific examples.)   

{\bf 3. Sodium and lithium clusters}

The magic numbers provided by the 3-dimensional $q$-deformed harmonic 
oscillator in Table 1 are compared to available experimental data for 
Na clusters \cite{Martin,Bjorn,Knight1,Peder} and Li clusters 
\cite{Brec,Brec2} in Table 4 (columns 2--7). Some preliminary results 
concerning Na clusters have already been given earlier in Ref. \cite{CPL302}. 
The following comments apply:

i) Only magic numbers up to 1500
are reported, since it is known that filling of electronic shells 
is expected to occur only up to this limit \cite{Martin}. For large 
clusters beyond this point it is known that magic numbers can be explained by
the completion of icosahedral or cuboctahedral shells of atoms \cite{Martin}. 

ii) Up to 600 particles there is consistency among the various experiments 
and between the experimental results in one hand and our findings in the 
other. 

iii) Beyond 600 particles the results of the four  experiments,
which report magic numbers in this region, are 
quite different. However, the results of all four  experiments are 
well accommodated by the present model. In addition, each magic number 
predicted by the model is supported by at least one experiment. 

In Table 4 the predictions of three simple theoretical models \cite{Mayer}
(non-deformed 3-dimensional harmonic oscillator (column 10), 
square well potential  
(column 9), rounded square well potential (intermediate between the 
previous two, column 8)~) are also reported for comparison. It is clear 
that the predictions of the non-deformed 3-dimensional harmonic oscillator
are 
in agreement with the experimental data only up to magic number 40, 
while the other two models give correctly a few more magic numbers (58, 
92, 138), although they already fail by predicting magic numbers at 68, 70, 
106, 112, 156, which are not observed.  

It should be noticed at this point that the first few magic numbers of 
alkali clusters (up to 92) can be correctly reproduced by the assumption 
of the formation of shells of atoms instead of shells of delocalized 
electrons \cite{Anagnos}, this assumption being applicable  under conditions 
not favoring delocalization of the valence electrons of alkali atoms. 

Comparisons among the present results, experimental data for Na clusters 
(by Martin {\it et al.} \cite{Martin} (column 2) and Pedersen {\it et al.}
\cite {Peder} (column 3)), experimental data for Li clusters
(Br\'echignac {\it et al.} \cite{Brec}
(column 4)~), and 
theoretical predictions more sophisticated than these reported in Table 4,
can be made in Table 5, where magic numbers predicted by various 
jellium model calculations (columns 5--8, 
\cite{Martin,Bjorn,Brack,Bulgac}), Woods--Saxon 
and wine bottle potentials (column 9, \cite{Nishi}), as well as by a 
classification scheme using the $3n+l$ pseudo quantum number 
(column 10, \cite{Martin}) are reported. The following observations can be 
made:

i) All magic numbers predicted by the 3-dimensional $q$-deformed harmonic 
oscillator are supported by at least one experiment, with no exception.

ii) Some of the jellium models, as well as the $3n+l$ classification scheme, 
predict magic numbers at 186, 540/542, which are not supported by 
experiment. Some jellium models also predict a magic number at 
748 or 758, again without support from experiment. The Woods--Saxon 
and wine bottle potentials of Ref. \cite{Nishi} predict a magic number at 
68, for which no experimental support exists. The present scheme 
avoids problems at these numbers. It should be noticed, however, 
that in the cases of 186 and 542 the energy gap following them 
in the present scheme is 
0.329 and 0.325 respectively (see Table 1), i.e. quite close to 
the threshold of 0.39 which we have considered as the minimum energy 
gap separating different shells. One could therefore qualitatively 
remark that 186 and 542 are ``built in'' the present scheme as
``secondary'' (not very pronounced) magic numbers.  

{\bf 4. Other alkali metals and noble metals} 

Experimental data for various alkali metal clusters [Li (\cite{Brec}, 
column 2), Na (\cite{Martin}, 
column 3), K (\cite{Knight2}, column 4), Rb (\cite{Bhaskar}, column 5), 
Cs (\cite{Bjorn,Goehlich}, column 6)] and 
noble metal clusters [Cu (\cite{Kata1}, column 7), Ag (\cite{Alam} in column 8
and \cite{Kata1} in column 9),
 Au (\cite{Kata1}, column 10)]  are reported in Table 6, along with the 
theoretical predictions of the 3-dimensional $q$-deformed harmonic oscillator 
given in Table 1. The following comments apply:

i) In the cases of Rb \cite{Bhaskar}, Cu \cite{Kata1}, Ag \cite{Kata1}, 
Au \cite{Kata1}, what is seen experimentally is cations 
of the type Rb$^+_N$, Cu$^+_N$, Ag$^+_N$, Au$^+_N$, which contain $N$ atoms 
each, but $N-1$ electrons. The magic numbers reported in  Table 6 are 
electron magic numbers in all cases. 

ii) All alkali metals and noble metals give the same magic numbers, at least 
within the ranges reported in the table. For most of these metals the 
range of experimentally determined magic numbers is rather limited, 
with Na \cite{Martin}, Cs \cite{Bjorn,Goehlich}, Li \cite{Brec}, and 
Ag \cite{Alam} being notable exceptions. 

iii) The magic numbers occuring in Na \cite{Martin}, Cs \cite{Bjorn,Goehlich},
Li \cite{Brec}, and Ag \cite{Alam}   are almost identical, and are 
described very well by the 3-dimensional $q$-deformed harmonic oscillator 
of Table 1. The limited data on K, Rb, Cu, Au, also agree with the magic 
numbers of Table 1.  

{\bf 5. Divalent metals of the IIB group}

For these metals the quantities determined experimentally \cite{Kata2} are 
numbers of atoms exhibiting ``magic''
behaviour. Each atom has two valence electrons, therefore the magic numbers 
of electrons are twice the magic numbers of atoms. The magic numbers of 
electrons for Zn and Cd clusters \cite{Kata2} are reported in Table 7
(in columns 4 and 5 respectively), along 
with the magic numbers predicted by the 3-dimensional $q$-deformed harmonic 
oscillator for two different parameter values (given in Tables 1 and 2 and 
reported in columns 1 and 2 respectively),
and the magic numbers given by a potential intermediate between the simple
harmonic oscillator and the square well potential (\cite{Kata2},
column 3). The following 
comments can be made:

i) The experimental magic numbers for Zn and Cd \cite{Kata2} are almost 
identical. Magic numbers reported in parentheses are ``secondary'' magic 
numbers, while the magic numbers without parentheses are the ``main'' ones, 
as indicated in \cite{deHeer}. 

ii) In column 1 of Table 7 magic numbers of the 3-dimensional $q$-deformed 
harmonic oscillator with $\tau=0.038$ and energy gaps larger than 0.26 are 
reported. Decreasing the energy gap $\delta$ considered as separating 
different shells from 0.39 (used in Table 1) to 0.26 (used in Table 7) has
as a
result that the numbers 70 and 106 become magic, in close agreement with 
the experimental data. Similar but even better results are gotten from 
the 3-dimensional $q$-deformed harmonic oscillator of Table 2, reported 
in column 2 of Table 7. This oscillator is characterized by $\tau =0.020$, 
while the energy gap $\delta$ between different shells is set equal to 0.20~. 
We observe that the second oscillator predicts an additional magic number 
at 112, in agreement with experiment, 
but otherwise gives the same results as the first one. 
We remark therefore that the general agreement between the results given 
by the 3-dimensional $q$-deformed harmonic oscillator and the experimental 
data is not sensitively dependent on the parameter value, but, in contrast, 
quite different parameter values ($\tau=0.038$, $\tau=0.020$) provide 
quite similar sets of magic numbers (at least in the region of relatively 
small magic numbers). 

iii) Both oscillators reproduce all the ``main'' magic numbers of Zn and Cd,
while the intermediate potential between the simple harmonic oscillator and 
the square well potential, reported in column 3,  reproduces all the
 ``main'' magic numbers except 108. 

{\bf 6. Trivalent metals of the III group} 

Magic numbers of electrons for the trivalent metals Al and In \cite{Persson}
are reported in Table 7 (in columns 7 and 8 respectively), along with the 
predictions of the 3-dimensional $q$-deformed harmonic oscillator of Table 3 
(column 6). The following comments can be made: 

i) It is known \cite{deHeer,Persson} that small magic numbers in 
clusters of Al and In cannot be explained by models based on the filling 
of electronic shells, because of symmetry breaking caused by the ionic 
lattice \cite{Persson}, while for large magic numbers this problem 
does not exist. 

ii) The 3-dimensional $q$-deformed harmonic oscillator of Table 3
provides the magic numbers reported in column 6 of Table 7. These magic 
numbers agree quite well with the experimental findings, with an 
exception in the region of small magic numbers, where the model
fails to reproduce 
the magic numbers 164 and 198, predicting only a magic number at 186.
In addition the oscillator predicts magic numbers at 398, 890, 1074, 
which are not seen in the experiment reported in column 7.  

{\bf 7. Potentials corresponding to the 3-dimensional $q$-deformed harmonic 
oscillator} 

As we have seen in the previous sections, 
the 3-dimensional $q$-deformed harmonic oscillator describes successfully 
the magic numbers of several metallic clusters. On the other hand, it is 
known that metallic clusters are successfully described by the Ekardt 
potentials \cite{Ekardt} 
(for which analytical expressions are lacking), which have been 
recently parametrized in terms of symmetrized Woods--Saxon and ``wine-bottle''
symmetrized Woods--Saxon potentials \cite{Kotsos,Kotsos2}
(for which analytical expressions are 
known). Therefore the following questions are created:

a)  Is it possible to determine some potentials
which, when put into the Schr\"odinger equation,  will provide approximately 
the same spectrum as the 3-dimensional $q$-deformed harmonic oscillator? 

b) If such potentials can be found, how do they compare with the symmetrized 
Woods--Saxon and ``wine-bottle'' symmetrized Woods--Saxon potentials? 

Question a) is a standard problem of inverse scattering \cite{Chadan}. 
Classical potentials giving approximately the same spectrum as the 
one-dimensional $q$-deformed harmonic oscillator have been determined 
either through use of standard perturbation theory \cite{BDKPRA}, or within 
the limits of the WKB approximation \cite{BDKJMP}. 

In what follows we are going to determine potentials giving approximately 
the same spectrum as the 3-dimensional $q$-deformed harmonic oscillator 
by using the method of Ref. \cite{BDKPRA}, i.e. perturbation theory. 
According to this method, a potential of the form 
\begin{equation}
V = V_0 +\kappa x^2 +\lambda x^4 + \mu x^6 + \xi x^8 +\cdots
\end{equation}
corresponds, in first-order perturbation theory and keeping terms up to 
$x^8$ only, to a spectrum 
$$ E=\epsilon_0+ \kappa + 3\lambda +15 \mu + 105 \xi + (2\kappa+6\lambda 
+40\mu + 280 \xi) n $$
\begin{equation}
+ (6\lambda+ 30\mu+350\xi) n^2 + 
(20\mu+140\xi) n^3 + 70\xi n^4.
\end{equation} 
The second term in Eq. (8) corresponds to the usual harmonic oscillator. 
For appropriate values of the numerical coefficients $\kappa$, $\lambda$, 
$\mu$, $\xi$, the rest of the terms can be considered as perturbations
to the harmonic oscillator. 

It is clear that this method can be applied in cases in which the spectrum 
under study depends on only one quantum number, the number of excitation 
quanta $n$. In the case of the 3-dimensional $q$-deformed harmonic 
oscillator (Eq. (3)), though, the spectrum depends on an additional quantum 
number, the angular momentum $l$. One way out is to determine 
a $l$-dependent equivalent potential, as it is done in several branches 
of physics \cite{Fiedeldey,Mackintosh}. 
In order to do this, for each possible value of $l$ ($l=n, 
n-2, n-4, \ldots 1$ or 0 (see Eq. (3)) one determines the energy as a
function 
of $n$ only  and then calculates the corresponding potential.  

In the case of the 3-dimensional $q$-deformed harmonic oscillator the energy 
spectrum $E_q(n,l)$ for the various possible values of the angular momentum 
$l$ ($l=n$, $n-2$, $n-4$, \dots  1 or 0) can be put in the form: 
\begin{equation}
E_q(n,n)= \hbar \omega_0 [n]_{q^2} \qquad {\rm for} \quad l=n, 
\end{equation}
\begin{equation}
E_q(n,n-2)= \hbar \omega_0 \left( q^2 [n-1]_{q^2} +q^{2n} \right) \qquad 
{\rm for} \quad l=n-2, 
\end{equation}
\begin{equation}
E_q(n,n-4)= \hbar \omega_0 \left( q^4 [n-2]_{q^2} + q^{2(n-1)} [2]_{q^2}
\right) \qquad {\rm for} \quad l=n-4, 
\end{equation}
\begin{equation}
E_q(n,n-6)= \hbar \omega_0 \left( q^6 [n-3]_{q^2} + q^{2(n-2)} [3]_{q^2}
\right) \qquad {\rm for} \quad l=n-6, 
\end{equation}
$$ \dots $$
$$E_q(n,3)= \hbar \omega_0 \left( q^8 [[n-3]]_{q^2} + q^{-4} [[5]]_{q^2} 
-q^{-2} [[3]]_{q^2} +1\right) $$ 
\begin{equation}
=E_q(n,0)-\hbar \omega_0 (q^6-1) (1+q^{-4})
\qquad {\rm for} \quad l=3,
\end{equation} 
\begin{equation}
E_q(n,2)=\hbar \omega_0 \left( q^6 [[n-2]]_{q^2} +q^{-2} [[3]]_{q^2} -1\right)
= E_q(n,0) -\hbar\omega_0 (q^4 -q^{-2})
\qquad {\rm for} \quad l=2, 
\end{equation}
\begin{equation}
E_q(n,1) =\hbar \omega_0 \left( q^4 [[n-1]]_{q^2} +1\right) 
= E_q(n,0) -\hbar \omega_0 (q^2-1)
\qquad {\rm for}\quad l=1, 
\end{equation}
\begin{equation}
E_q(n,0)=\hbar \omega_0 q^2 [[n]]_{q^2} \qquad {\rm for} \quad l=0,
\end{equation}
where by $[n]_q$ are denoted the $q$-numbers of Eq. (2), which are 
symmetric under the exchange $q\leftrightarrow q^{-1}$, while by 
$[[n]]_q$ are denoted the $q$-numbers 
\begin{equation}
[[n]]_q = {q^n -1\over q-1},
\end{equation}
which are not symmetric under the exchange $q \leftrightarrow q^{-1}$. 
For all of these equations it is clear that they reduce to the classical 
expression $E(n)=\hbar \omega_0 n $ in the limit $q\rightarrow 1$. 

We then consider the Taylor expansions for these energy 
expressions. By comparing them to Eq. (9) and equating the coefficients 
of the various powers of $n$ (up to $n^4$) we determine in each case the 
coefficients $\kappa$, $\lambda$, $\mu$, $\xi$. Substituting these 
coefficients in Eq. (8) we determine for each case the corresponding 
potential, keeping terms up to $\tau^4$ (where $q=e^\tau$). 
The first few cases are:
$$ {V(x)_{l=n}\over\hbar\omega_0}= 
-\left({1\over 2} -{\tau^2\over 2}+{4\tau^4 \over 15}\right) 
+ \left( {1\over 2} -{\tau^2 \over 2} + 
{4 \tau^4\over 15}\right) x^2 $$ 
\begin{equation}
-\left( {\tau^2\over 6}-{\tau^4\over 9}\right) x^4 +\left( {\tau^2\over 30}
-{\tau^4\over 45}\right) x^6 \qquad {\rm for} \quad l=n,
\end{equation}
$${V(x)_{l=n-2}\over \hbar\omega_0} 
=-\left({1\over 2} +4\tau+ {7\tau^2\over 2} +{8\tau^3\over 3} +
{8 \tau^4\over 5} \right) + \left( {1\over 2}+2\tau +{3\tau^2\over 2} 
+{10\tau^3\over 3} + {44\tau^4\over 15}\right) x^2 $$
\begin{equation}
-\left( {\tau^2\over 6} +{4\tau^3 \over 3} +{11 \tau^4 \over 9}\right) x^4 
+\left( {\tau^2\over 30} + {2\tau^3\over 15} +{\tau^4 \over 9}\right) x^6
\qquad {\rm for} \quad l=n-2,
\end{equation}
$$\dots$$
$$ {V(x)_{l=0}\over \hbar \omega_0} = 
-\left({1\over 2}+{\tau\over 2} -{\tau^3\over 6} -{\tau^4\over 15}\right) 
+\left( {1\over 2} -{\tau^2 \over 2} 
+{4 \tau^4 \over 15}\right) x^2 + \left( {\tau\over 6} -{2\tau^3\over 9}
-{\tau^4 \over 6}\right) x^4 $$
\begin{equation}
+ \left( {\tau^2 \over 30} -{\tau^4\over 45}
\right) x^6 +\left( {\tau^3 \over 210} +{\tau^4 \over 210}\right) x^8
\qquad {\rm for} \quad l=0,
\end{equation} 
$$ V(x)_{l=1} = V(x)_{l=0} -\hbar \omega_0 (q^2-1) $$
\begin{equation}
\simeq V(x)_{l=0} -\hbar \omega_0 \left( 2 \tau + 2 \tau^2 +{4\tau^3 \over 3} 
+{2\tau^4 \over 3}\right) 
\qquad {\rm for} \quad l=1,
\end{equation}  
$$ V(x)_{l=2}= V(x)_{l=0} -\hbar \omega_0 (q^4 -q^{-2}) $$
\begin{equation}
\simeq V(x)_{l=0}-\hbar \omega_0 (6\tau+6\tau^2 +12 \tau^3 +10 \tau^4)
\qquad {\rm for} \quad l=2, 
\end{equation}
$$V(x)_{l=3} = V(x)_{l=0} -\hbar \omega_0 (q^6-1) (1+q^{-4}) $$
\begin{equation}
\simeq V(x)_{l=0}-\hbar \omega_0 (12\tau+12 \tau^2 +48 \tau^3 +44 \tau^4) 
\qquad {\rm for} \quad l=3,
\end{equation}
$$ \dots$$

We remark that for small values of $\tau$, like the ones occuring in the 
previous sections, the potentials occuring for $l=n$ and $l=n-2$  are of 
the form 
\begin{equation}
V(x) = V_0 + a x^2 -b x^4 + c x^6, 
\end{equation}
with $a,b,c>0$, while the potentials occuring for $l=0$, 1, 2, 3 are of the 
form 
\begin{equation}
V(x)= V_0 + a x^2 + bx^4 + c x^6 + d x^8,
\end{equation}
with $a,b,c,d>0$.  

It is instructive at this point to compare these potentials with the 
symmetrized Woods--Saxon potential
\begin{equation}
V_{SWS}(r)=-V_0{\sinh(R/a) \over \cosh(r/a)+\cosh(R/a)}, \qquad 
0\leq r \leq \infty,
\end{equation} 
and the ``wine-bottle'' symmetrized Woods--Saxon potential 
\begin{equation}
V_{WB}(r)= -V_0 \left( 1+{w r^2 \over R^2}\right) {\sinh(R/a) \over 
\cosh(r/a) + \cosh(R/a)}, \qquad 0\leq r \leq \infty,
\end{equation}  
which have been used \cite{Kotsos,Kotsos2} for parametrizing 
the Ekardt potentials \cite{Ekardt}. 
In order to facilitate the comparisons, we consider the Taylor expansions 
of these potentials 
$$ {V_{SWS}(r) \over V_0 \sinh(R/a)} = -{1\over 1+\cosh(R/a)} +{1 \over 
2(1+\cosh(R/a))^2} {r^2\over a^2}  $$
\begin{equation}
- {5-\cosh(R/a) \over 24(1+\cosh(R/a))^3} {r^4\over a^4} 
+ {(\cosh(R/a))^2-28\cosh(R/a)+61    \over 720 (1+\cosh(R/a))^4}
{r^6\over a^6},
\end{equation}
$${V_{WB}(r)\over V_0 \sinh(R/a)} = -{1\over 1+\cosh(R/a)} +\left(
{1\over 2(1+\cosh(R/a))^2 } -{w\over 1+\cosh(R/a)} {a^2\over R^2} \right)
{r^2\over a^2} $$
$$-\left( {5-\cosh(R/a)\over 24(1+\cosh(R/a))^3} 
-{w\over 2(1+\cosh(R/a))^2} {a^2 \over R^2} \right) {r^4\over a^4} $$
\begin{equation}
+\left( {(\cosh(R/a))^2-28\cosh(R/a)+61\over 720 (1+\cosh(R/a))^4} 
-{w(5-\cosh(R/a))\over 24(1+\cosh(R/a))^3} {a^2\over R^2}\right) 
{r^6\over a^6}.
\end{equation} 

The following comments are now in place:

i) The Taylor expansions of the symmetrized Woods--Saxon 
and the ``wine-bottle'' symmetrized Woods--Saxon potentials,
which have been used for fitting the Ekardt potentials used for the 
description of metallic clusters, have the same form 
as the potentials corresponding to the 3-dimensional $q$-deformed harmonic 
oscillator, i.e. they contain all the even powers of the relevant variable
(and no odd powers).  It is therefore not surprising that the 3-dimensional 
$q$-deformed harmonic oscillator gives a good description of the magic 
numbers of metallic clusters. 

ii) The potentials obtained through the use of perturbation theory are valid 
near the origin ($x=0$) and for relatively low values of $n$. They do not give
information about the shape of the 
potential near its edges, or for very large values of $n$.
The determination of potentials which will be accurate near their edges 
remains an open problem. It should also be examined if these potentials
possess any deeper relation to the quantum algebraic symmetry characterizing 
the 3-dimensional $q$-deformed harmonic oscillator. For example, one could 
check if these potentials are related to the generators of the relevant 
quantum algebra. The existence of such a relation also remains an open 
problem. 

iii) For very large values of $n$, the spectrum gets an exponential form.
For example, Eq. (10) becomes (for $\tau>0$) 
\begin{equation}
E_q(n,n)= \hbar \omega_0 {e^{2\tau n} -e^{-2\tau n}\over e^{2\tau}
-e^{-2\tau} } \simeq \hbar \omega_0 {e^{2\tau n} \over e^{2\tau} 
-e^{-2\tau}}.
\end{equation} 
Potentials with exponential spectra have been considered in Ref. 
\cite{Spiridonov}, but also in this case only the form of the potential 
near the origin could be determined. 

iv) Focusing potentials leading to $3n+l$ degeneracy of the energy levels
(which has been found to describe reasonably well the magic numbers of 
alkali clusters \cite{Martin}) have been determined in Ref. \cite{Ostrovsky}.
They have the form
\begin{equation}
U_3(r) = -{2v\over R^4} { (r/R)^4 \over [(r/R)^6+1]^2},
\end{equation}
\begin{equation}
V_{\tilde 3}(r) = E - {2 L_m^2 \over m R_m^2} { (r/R_m)^4 \over 
[(r/R_m)^6+1]^2}.
\end{equation}
Both of them are of the form
\begin{equation}
V(x)= E -A {x^4 \over (x^6+1)^2},
\end{equation}
which corresponds to a Taylor expansion of the form
\begin{equation}
V(x)= E-A (x^4 -2 x^{10} + 3 x^{16} +\cdots).
\end{equation} 
We remark that this Taylor expansion bears no similarity to the Taylor 
expansions of the symmetrized Woods--Saxon and ``wine-bottle'' symmetrized 
Woods--Saxon potentials, since it contains only some of the even powers 
of the relevant variable and not all of them.  
Indeed, these focusing potentials are known to 
exhibit the ``wine-bottle'' feature strongly exaggerated \cite{Ostrovsky},
lacking in parallel the flat bottom characterizing the Woods--Saxon and 
Ekardt potentials. The potential $U_3(r)$
has, however, the major advantage that it reproduces quite well 
the edge behaviour of the Ekardt potentials \cite{Ostrovsky}. 

{\bf 8. Discussion} 

The following general remarks can now be made:

i) From the results reported above it is quite clear that the 
3-dimensional $q$-deformed harmonic oscillator describes very well 
the magic numbers of alkali metal clusters and noble metal clusters 
in all regions, using only one free parameter ($q=e^{\tau}$ with 
$\tau=0.038$). It also provides an accurate description of the 
``main'' magic numbers of clusters of divalent group IIB metals,
either with the same parameter value ($\tau=0.038$) or with 
a different one ($\tau=0.020$). In addition it gives a satisfactory 
description of the magic numbers of clusters of trivalent group III metals
with a different parameter value ($\tau=0.050$). 

ii) It is quite remarkable that the 3-dimensional $q$-deformed harmonic 
oscillator reproduces long sequences of
magic numbers (Na, Cs, Li, Ag) at least as accurately as other,
more sophisticated, models by using only one free parameter ($q=e^{\tau}$). 
(It should not be forgotten at this point that these other models have 
deep physical roots, while the present approach is based on symmetry 
arguments, which are justified a posteriori by their successful predictions.)
Once the parameter is fixed, the whole spectrum is fixed and no further 
manipulations can be made, the choice of the energy gap $\delta$ being 
the only exception. However, the choice of the value of the energy gap
$\delta$ does not influence the order of the energy levels, but 
it is just deciding which energy separations will be considered as 
corresponding to main magic numbers and which not. 
The successful prediction of the magic numbers 
can be considered as evidence that the 3-dimensional $q$-deformed 
harmonic oscillator owns a symmetry (the u$_q$(3) $\supset$ so$_q$(3)
symmetry) appropriate for the description of the physical systems under 
study. 

iii) As we have already mentioned, 
it has been remarked \cite{Martin} that if $n$ is the number of nodes 
in the solution of the radial Schr\"odinger equation and $l$ is the 
angular momentum quantum number, then the degeneracy of energy levels of 
the hydrogen atom characterized by the same $n+l$ is due to the so(4) 
symmetry of this system, while the degeneracy of energy levels of the 
spherical harmonic oscillator (i.e. of the 3-dimensional isotropic 
harmonic oscillator) characterized by the same $2n+l$ 
is due to the su(3) symmetry of this system. $3n+l$ has been used 
\cite{Martin} to approximate the magic numbers of alkali metal clusters
with some success, and focusing potentials characterized by this degeneracy 
have been determined \cite{Ostrovsky}, but no relevant Lie symmetry could be 
determined \cite{Ostrovsky,Koch}. In view
of the present findings the lack of Lie symmetry related to $3n+l$ is quite 
clear: the symmetry of the system appears to be a quantum algebraic 
symmetry (u$_q$(3)), which is a nonlinear extension of the Lie 
symmetry u(3). 

iv) The ability of the 3-dimensional $q$-deformed harmonic oscillator 
to reproduce correctly the magic numbers of several metal clusters 
is not coming as a surprise, if one considers potentials giving approximately 
(within the limits of perturbation theory) the same spectrum as this 
oscillator. The Taylor expansions of these potentials have the same form 
as the Taylor expansions of the symmetrized Woods--Saxon and ``wine-bottle''
symmetrized Woods--Saxon potentials, which successfully fit 
\cite{Kotsos,Kotsos2} the Ekardt 
potentials \cite{Ekardt}, which characterize the structure of metal
clusters.  

In summary, we have shown that the 3-dimensional $q$-deformed harmonic 
oscillator with u$_q$(3) $\supset$ so$_q$(3) symmetry correctly 
predicts all experimentally observed magic numbers of alkali metal clusters
and of noble metal clusters   
up to 1500, which is the expected limit of validity for theories based on 
the filling of electronic shells. In addition it gives a good description 
of the ``main'' magic numbers of group IIB (divalent) metal clusters, 
as well as a satisfactory description of group III (trivalent) metal
clusters. 
This indicates that u$_q$(3), which 
is a nonlinear deformation of the u(3) symmetry of the spherical
(3-dimensional isotropic) harmonic oscillator, is a good candidate for 
being the symmetry of systems of several  metal clusters.  
Furthermore, the Taylor expansions of potentials giving approximately the
same 
spectrum as the 3-dimensional $q$-deformed harmonic oscillator are found 
to have the same form as the Taylor expansions of the symmetrized 
Woods--Saxon and ``wine-bottle'' symmetrized Woods--Saxon potentials,
which sucsessfully fit the Ekardt potentials underlying the structure 
of metal clusters. Naturally, these Taylor expansions are valid near the 
origin. The determination of potentials which will be accurate near 
their edges remains an open problem. It is also an open problem the existence 
of any deeper relation between these potentials and the 
quantum algebra characterizing the 3-dimensional $q$-deformed harmonic 
oscillator, as, for example, some relation between these potentials and the 
generators of the quantum algebra.  

{\bf Acknowledgements}

One of the authors (PPR) acknowledges support from the Bulgarian Ministry 
of Science and Education under contracts $\Phi$-415 and $\Phi$-547, 
while another author (NK) has been supported by the Greek General 
Secretariat of Research and Technology under Contract No. PENED95/1981.

\newpage
%%%%%%%%%%%%%%%%%%%%%%% Table 1 %%%%%%%%%%%%%%%%%%%%%%%%%%%%%%%%%
\oddsidemargin -0.25cm\evensidemargin -0.25cm 
\topmargin -3.0cm 
\textwidth 16.3cm
\textheight 25.3cm

\baselineskip=22pt plus 1pt minus 1pt

\parindent=0pt
%%%%%%%%%%%%%%%%%%%%%%%%%%%%%%%%%%%%%%%%%%%%%%%%%%%%%%%%%%%%%%%%%%%%%%
%%%%%%%%%%%%%%%%%%% Table 1 %%%%%%%%%%%%%%%%%%%%%%%%%%%%%%%%%%%%%%%%

Table 1: 
Energy spectrum, $E_q(n,l)$,  of the 3-dimensional $q$-deformed 
harmonic oscillator (Eq. (3)), for $\hbar \omega_0 =1$ and 
$q=e^\tau$ with $\tau = 0.038$. Each level is characterized by $n$ 
(the number of vibrational quanta) and  $l$ (the angular momentum).
$2(2l+1)$ represents the number of particles each level can accommodate,
while under ``total'' the total number of particles up to and including 
this level is given. Magic numbers, reported in boldface, correspond to 
energy gaps larger than $\delta =0.39$, reported between the relevant 
couples of energy levels.

\bigskip
\vfill\eject

\begin{table}

\caption{ }
\bigskip
\begin{tabular}{r r r r r || r r r r r}
\hline
$n$ & $l$ & $E_q(n,l)$ & $2(2l+1)$ & total & $n$ & $l$ & $E_q(n,l)$ & 
$2(2l+1)$ & total \\
\hline
 0&  0&  0.000 &  2  &  {\bf 2}  &    9&  5&  12.215 & 22&  462 \\
  &   &  1.000 &     &           &   11& 11&  12.315 & 46&  508 \\
 1&  1&  1.000 &  6  &  {\bf 8}  &   10&  8&  12.614 & 34&  542 \\
  &   &  1.006 &     &           &    9&  3&  12.939 & 14&  {\bf 556} \\
 2&  2&  2.006 & 10  & 18  &     &   &   0.397 &   &      \\
 2&  0&  2.243 &  2  & {\bf 20}  &    9&  1&  13.336 &  6&  562 \\
  &   &  0.780 &     &           &   12& 12&  13.721 & 50&  612 \\
 3&  3&  3.023 & 14  & {\bf 34}  &   10&  6&  13.863 & 26&  638 \\
  &   &  0.397 &     &           &   11&  9&  14.154 & 38&  {\bf 676} \\
 3&  1&  3.420 &  6  & {\bf 40}  &     &   &   0.603 &   &      \\
  &   &  0.638 &     &           &   10&  4&  14.757 & 18&  {\bf 694} \\
 4&  4&  4.058 & 18  & {\bf 58}  &     &   &   0.449 &   &      \\
  &   &  0.559 &     &           &   13& 13&  15.206 & 54&  748 \\
 4&  2&  4.617 & 10  & 68        &   10&  2&  15.316 & 10&  758 \\
 4&  0&  4.854 &  2  & 70        &   10&  0&  15.554 &  2&  760 \\
 5&  5&  5.116 & 22  & {\bf 92}  &   11&  7&  15.592 & 30&  790 \\
  &   &  0.724 &     &           &   12& 10&  15.777 & 42&  {\bf 832} \\
 5&  3&  5.841 & 14  &106        &     &   &   0.884 &   &      \\
 6&  6&  6.204 & 26  &132        &   11&  5&  16.660 & 22&  854 \\
 5&  1&  6.238 &  6  &{\bf 138}  &   14& 14&  16.779 & 58&  {\bf 912} \\
  &   &  0.860 &     &           &     &   &   0.606 &   &      \\
 6&  4&  7.098 & 18  &156        &   11&  3&  17.385 & 14&  926 \\
 7&  7&  7.328 & 30  &186        &   12&  8&  17.410 & 34&  960 \\
 6&  2&  7.657 & 10  &196        &   13& 11&  17.490 & 46& 1006 \\
 6&  0&  7.895 &  2  &{\bf 198}  &   11&  1&  17.782 &  6& {\bf 1012} \\
  &   &  0.502 &     &           &     &   &   0.667 &   &      \\
 7&  5&  8.396 & 22  &220        &   15& 15&  18.449 & 62& 1074 \\
 8&  8&  8.494 & 34  &{\bf 254}  &   12&  6&  18.660 & 26& {\bf 1100} \\
  &   &  0.627 &     &           &     &   &   0.645 &   &      \\
 7&  3&  9.121 & 14  &{\bf 268}  &   14& 12&  19.305 & 50& 1150 \\
  &   &  0.397 &     &           &   13&  9&  19.330 & 38& 1188 \\
 7&  1&  9.518 &  6  &274        &   12&  4&  19.554 & 18& {\bf 1206} \\
 9&  9&  9.709 & 38  &312        &     &   &   0.559 &   &      \\
 8&  6&  9.743 & 26  &{\bf 338}  &   12&  2&  20.113 & 10& 1216 \\
  &   &  0.894 &     &           &   16& 16&  20.226 & 66& 1282 \\
 8&  4& 10.637 & 18  &356        &   12&  0&  20.350 &  2& {\bf 1284} \\
10& 10& 10.980 & 42  &398        &     &   &   0.417 &   &      \\
 9&  7& 11.146 & 30  &428        &   13&  7&  20.767 & 30& {\bf 1314} \\
 8&  2& 11.196 & 10  &438        &     &   &   0.464 &   &      \\
 8&  0& 11.434 &  2  &{\bf 440}  &   15& 13&  21.231 & 54& 1368 \\
  &   &  0.781 &     &           &   14& 10&  21.360 & 42& {\bf 1410} \\
  &   &        &     &           &     &   &   0.475 &   &      \\
  &   &        &     &           &   13&  5&  21.835 & 22& 1432 \\
  &   &        &     &           &   17& 17&  22.119 & 70& {\bf 1502} \\
  &   &        &     &           &     &   &   0.441 &   &      \\
  &   &        &     &           &   13&  3&  22.560 & 14& 1516 \\
\hline
\end{tabular}
\end{table}

\newpage 
%%%%%%%%%%%%%%%%%%%%% Table 2 %%%%%%%%%%%%%%%%%%%%%%%%%%%%%%%%%%%%%%
\setcounter{table}{1}

%%%%%%%%%%%%%%%%%%%%%%%%%%%%%%%%%%%%%%%%%%%%%%%%%%%%%%%%%%%%%%%%%%%%%%
%%%%%%%%%%%%%%%%%%% Table 2 %%%%%%%%%%%%%%%%%%%%%%%%%%%%%%%%%%%%%%%%

\begin{table}
\caption{ 
Same as Table 1, but with $\hbar \omega_0 =1$ and $q=e^\tau$ with 
$\tau = 0.020$. The energy gap separating different shells has been
taken to be $\delta =0.20$~. }

\bigskip
\begin{tabular}{r r r r r || r r r r r}
\hline
$n$ & $l$ & $E_q(n,l)$ & $2(2l+1)$ & total & $n$ & $l$ & $E_q(n,l)$ & 
$2(2l+1)$ & total \\
\hline
 0&  0&  0.000 &  2  &  {\bf 2}  &    5&  5&   5.032 & 22& {\bf 92}\\
  &   &  1.000 &     &           &     &   &   0.369 &   &      \\
 1&  1&  1.000 &  6  &  {\bf 8}  &    5&  3&   5.401 & 14& {\bf 106} \\
  &   &  1.002 &     &           &     &   &   0.205 &   &            \\
 2&  2&  2.002 & 10  & 18        &    5&  1&   5.606 &  6&  {\bf 112} \\
 2&  0&  2.124 &  2  & {\bf 20}  &     &   &   0.450 &   &      \\
  &   &  0.882 &     &           &    6&  6&   6.056 & 26&  {\bf 138} \\
 3&  3&  3.006 & 14  & {\bf 34}  &     &   &   0.453 &   &      \\
  &   &  0.205 &     &           &    6&  4&   6.509 & 18&  {\bf 156} \\
 3&  1&  3.211 &  6  & {\bf 40}  &     &   &   0.286 &   &      \\
  &   &  0.805 &     &           &    6&  2&   6.795 & 10&       166  \\
 4&  4&  4.016 & 18  & {\bf 58}  &    6&  0&   6.918 &  2&  168 \\
  &   &  0.287 &     &           &    7&  7&   7.090 & 30&  {\bf 198}\\
 4&  2&  4.303 & 10  & 68        &     &   &   0.536 &   &      \\
 4&  0&  4.425 &  2  & {\bf 70}  &    7&  5&   7.626 & 22&  {\bf 220} \\
  &   &  0.607 &     &           &     &   &   0.369 &   &      \\
\hline
\end{tabular}
\end{table}

\newpage
%%%%%%%%%%%%%%%%%%%%%%% Table 3 %%%%%%%%%%%%%%%%%%%%%%%%%%%%%%%%%

\setcounter{table}{2}

%%%%%%%%%%%%%%%%%%%%%%%%%%%%%%%%%%%%%%%%%%%%%%%%%%%%%%%%%%%%%%%%%%%%%%
%%%%%%%%%%%%%%%%%%% Table 3 %%%%%%%%%%%%%%%%%%%%%%%%%%%%%%%%%%%%%%%%

\begin{table}
%Table 3: 
\caption{
Same as Table 1, but with $\hbar \omega_0 =1$ and $q=e^\tau$ with 
$\tau = 0.050$. The energy gap separating different shells has been
 taken to be $\delta=0.38$~.  }

%\begin{table}

%\caption{ }
\bigskip
\begin{tabular}{r r r r r || r r r r r}
\hline
$n$ & $l$ & $E_q(n,l)$ & $2(2l+1)$ & total & $n$ & $l$ & $E_q(n,l)$ & 
$2(2l+1)$ & total \\
\hline
 0&  0&  0.000 &  2  &  {\bf 2}  &   11& 11&  13.334 & 46&{\bf 486} \\
  &   &  1.000 &     &           &     &   &   0.389 &   &      \\
 1&  1&  1.000 &  6  &  {\bf 8}  &    9&  5&  13.723 & 22&  508 \\
  &   &  1.010 &     &           &   10&  8&  14.044 & 34&  {\bf 542} \\
 2&  2&  2.010 & 10  & 18        &     &   &   0.658 &   &      \\
 2&  0&  2.327 &  2  & {\bf 20}  &    9&  3&  14.702 & 14&  556 \\
  &   &  0.713 &     &           &   12& 12&  15.069 & 50&  606 \\
 3&  3&  3.040 & 14  & {\bf 34}  &    9&  1&  15.233 &  6& {\bf 612}\\
  &   &  0.531 &     &           &     &   &   0.540 &   &            \\
 3&  1&  3.571 &  6  & {\bf 40}  &   10&  6&  15.773 & 26& 638  \\
  &   &  0.530 &     &           &   11&  9&  15.971 & 38&  {\bf 676} \\
 4&  4&  4.101 & 18  & {\bf 58}  &     &   &   0.985 &   &      \\
  &   &  0.751 &     &           &   13& 13&  16.956 & 54&  730 \\
 4&  2&  4.852 & 10  & 68        &   10&  4&  16.989 & 18&{\bf 748} \\  
 4&  0&  5.168 &  2  & 70        &     &   &   0.751 &   &      \\
 5&  5&  5.202 & 22  & {\bf 92}  &   10&  2&  17.740 & 10&  758 \\
  &   &  0.979 &     &           &   11&  7&  17.981 & 30&  788       \\
 5&  3&  6.181 & 14  &106        &   10&  0&  18.056 &  2&  790 \\
 6&  6&  6.356 & 26  &132        &   12& 10&  18.057 & 42&{\bf 832}\\
 5&  1&  6.712 &  6  &{\bf 138}  &     &   &   0.954 &   &            \\
  &   &  0.860 &     &           &   14& 14&  19.011 & 58& {\bf 890}\\
 6&  4&  7.572 & 18  &156        &     &   &   0.435 &   &      \\
 7&  7&  7.573 & 30  &{\bf 186}  &   11&  5&  19.446 & 22&{\bf 912}\\
  &   &  0.750 &     &           &     &   &   0.878 &   &       \\
 6&  2&  8.323 & 10  &196        &   13& 11&  20.324 & 46&  958 \\
 6&  0&  8.639 &  2  &     198   &   12&  8&  20.368 & 34&  992       \\
 8&  8&  8.866 & 34  &232        &   11&  3&  20.424 & 14& {\bf 1006}\\
 7&  5&  9.038 & 22  &{\bf 254}  &     &   &   0.531 &   &      \\
  &   &  0.979 &     &           &   11&  1&  20.955 &  6& 1012       \\
 7&  3& 10.017 & 14  &     268   &   15& 15&  21.257 & 62& {\bf 1074}\\
 9&  9& 10.248 & 38  & 306       &     &   &   0.840 &   &      \\
 7&  1& 10.548 &  6  & 312       &   12&  6&  22.097 & 26& {\bf 1100} \\
 8&  6& 10.595 & 26  &{\bf 338}  &     &   &   0.697 &   &      \\
  &   &  1.137 &     &           &   14& 12&  22.794 & 50& 1150 \\
10& 10& 11.732 & 42  & 380       &   13&  9&  22.960 & 38& 1188 \\
 8&  4& 11.811 & 18  &{\bf 398}  &   12&  4&  23.313 & 18& {\bf 1206} \\
  &   &  0.447 &     &           &     &   &   0.403 &   &      \\
 9&  7& 12.258 & 30  &428        &   16& 16&  23.716 & 66&      1272  \\
 8&  2& 12.562 & 10  &438        &   12&  2&  24.064 & 10& 1282 \\
 8&  0& 12.878 &  2  &{\bf 440}  &   12&  0&  24.381 &  2& {\bf 1284}\\
  &   &  0.456 &     &           &     &   &   0.589 &   &            \\
  &   &        &     &           &   13&  7&  24.970 & 30& {\bf 1314}\\
\hline
\end{tabular}
\end{table}

\newpage
%%%%%%%%%%%%%%%%%%%%%%%% Table 4 %%%%%%%%%%%%%%%%%%%%%%%%%%%%%%%%%%%
\setcounter{table}{3}

%%%%%%%%%%%%%%%%%%%%%%%%%%%%%%%%%%%%%%%%%%%%%%%%%%%%%%%%%%%%%%%%%%%%%%
%%%%%%%%%%%%%%%%%%% Table 4 %%%%%%%%%%%%%%%%%%%%%%%%%%%%%%%%%%%%%%%%

\begin{table}

\caption{ Magic numbers provided by the 3-dimensional $q$-deformed harmonic 
oscillator (Table 1), reported in column 1, are compared to the experimental 
data for Na clusters by Martin {\it et al.} [6] (column 2), 
Bj{\o}rnholm {\it et al.} [7] (column 3), 
Knight {\it et al.} [8] (column 4), and 
Pedersen {\it et al.} [9] (column 5), as well as to the experimental data 
for Li clusters by Br\'echignac {\it et al.} ([10] in column 6, 
[11] in column 7).
The magic numbers provided [5] by the (non-deformed) 3-dimensional harmonic 
oscillator (column 10), the square well potential (column 9) and a rounded
square well potential intermediate between the previous two (column 8) 
are also shown for comparison. See text for discussion. }

\bigskip

\begin{tabular}{c c c c c c c c c c}
\hline
 th.&   exp.  & exp. &  exp.&  exp.& exp. & exp.  & th.  &  th. &  th.   \\
present& Na  &  Na & Na &  Na  &  Li & Li & int. & sq. well & h. osc.\\
%  our    Martin     Bjorn        Knight  Peder   Brec int    well        HO
 Tab. 1  & Ref.[6] &Ref.[7]& Ref.[8]& Ref.[9]& Ref.[10] & Ref.[11] &
Ref.[5] & Ref.[5] & Ref.[5] \\  
\hline
    2  &    2      &   2     & 2&      &      &   2&     2   &   2     &
2 \\
    8  &    8      &   8     & 8&      &      &   8&     8   &   8     &
8 \\
  (18) &   18      &         &  &      &      &    &    18   &  18     &
 \\
   20  &   20      &  20     &20&      &      &  20&    20   &  20     &
20 \\
   34  &   34      &         &  &      &      &    &    34   &  34     &
 \\
   40  &   40      &  40     &40&   40 &      &  40&    40   &  40     &
40 \\
   58  &   58      &  58     &58&   58 &      &  58&    58   &  58     &
 \\
       &           &         &  &      &      &    &  68,70  &  68     &
70 \\
   92  &  90,92    &  92     &92&   92 &   93 &  92&    92   & 90,92   &
 \\
       &           &         &  &      &      &    &  106,112&  106    &
112 \\
  138  &  138      & 138     &  &  138 &  134 & 138&    138  &  132,138&
 \\
  198  &  198$\pm$2 & 196     & &  198 &  191 & 198&    156  &   156   &
168 \\
  254  &           & 260$\pm$4& &      &      & 258&         &         &
 \\
  268  &  263$\pm$5 &         & &  264 &  262 &    &         &         &
 \\
  338  & 341$\pm$5 & 344$\pm$4& &  344 &  342 & 336&         &         &
 \\
  440  & 443$\pm$5 & 440$\pm$2& &  442 &  442 & 440&         &         &
 \\
  556  & 557$\pm$5 & 558$\pm$8& &  554 &  552 & 546&         &         &
 \\
  676  &           &         &  &  680 &      &    &         &         &
 \\
  694  &  700$\pm$15&         & &      &  695 & 710&         &         &
 \\
  832  &  840$\pm$15&         & &  800 &  822 & 820&         &         &
 \\
  912  &           &         &  &      &  902 &    &         &         &
 \\
 1012  & 1040$\pm$20&         & &  970 & 1025 &1065&         &         &
 \\
 1100  &           &         &  & 1120 &      &    &         &         &
 \\
 1206  & 1220$\pm$20&         & &      &      &    &         &         &
 \\
 1284  &           &         &  &      & 1297 &1270&         &         &
 \\
 1314  &           &         &  & 1310 &      &    &         &         &
 \\
 1410  &  1430$\pm$20&        & &      &      &    &         &         &
 \\
 1502  &           &         &  & 1500 &      &1510&         &         &
 \\
\hline
\end{tabular}
\end{table}

\newpage
%%%%%%%%%%%%%%%%%%%%%%%%%% Table 5 %%%%%%%%%%%%%%%%%%%%%%%%%%%%%%%%%%%
\setcounter{table}{4}
%%%%%%%%%%%%%%%%%%% Table 5 %%%%%%%%%%%%%%%%%%%%%%%%%%%%%%%%%%%%%%%%

\begin{table}

\caption{Magic numbers provided by the 3-dimensional $q$-deformed harmonic 
oscillator (Table 1), reported in column 1, are compared to the experimental 
data for Na clusters by  Martin {\it et al.} [6] (column 2), and 
Pedersen {\it et al.} [9] (column 3), as well as to the experimental data 
for Li clusters by  
Br\'echignac {\it et al.} [10] (column 4) and 
to the theoretical predictions of various jellium model 
calculations reported by Martin {\it et al.} [6] (column 5), Bj{\o}rnholm 
{\it et al.} [7] (column 6), Brack [2] (column 7), Bulgac and Lewenkopf
[42] (column 8), the theoretical predictions of Woods--Saxon and 
``wine bottle'' potentials reported by Nishioka {\it et al.} [43] (column 9),
as well as to the magic numbers predicted by the  classification scheme 
using the $3n+l$ pseudo quantum number, reported by Martin {\it et al.}
[6] (column 10). See text for discussion.  
}

\bigskip

\begin{tabular}{c c c c c c c c c c}
 \hline
 th.    &exp. &exp. &exp. &th.     &  th.  &  th.    & th.   &  th. & th.   \\
present & Na   & Na  & Li  &jell.   & jell. &  jell.  & jell. & WS  &$3n+l$ \\
Tab. 1  & Ref.[6] &Ref.[9]& Ref.[10]& Ref.[6]& Ref.[7] & Ref.[2] 
& Ref.[42]& Ref.[43]  & Ref.[6] \\  
\hline
%          exp   exp   exp jell        jell     jell          
%  our    Martin Ped  Brec Martin      Bjorn    Brack Bulgac  Nishi    3n+l 
  2 &    2     &     &     &  2     &    2 &     2   &       &     2  &
2 \\
  8 &    8     &     &     &  8     &    8 &     8   &       &     8  &
8 \\
(18)&   18     &     &     & 18     &   18 &         &       &        &
18 \\
 20 &   20     &     &     &(20)    &   20 &    20   &       &    20  &
 \\
 34 &   34     &     &     & 34     &   34 &    34   &  34   &        &
34 \\
 40 &   40     &  40 &     &(40)    &   40 &         &       &    40  &
 \\
 58 &   58     &  58 &     & 58     &   58 &    58   &  58   &    58  &
58 \\
    &          &     &     &        &      &         &       &    68  &
 \\
 92 &  90,92   &  92 &  93 & 92     &   92 &    92   &  92   &    92  &
90 \\
138 &  138     & 138 & 134 &134     &  138 &   138   &  138  &   138  &
132 \\
    &          &     &     &186     &  186 &   186   &  186  &        &
186 \\
198 & 198$\pm$2& 198 & 191 &(196)   &  196 &         &       &   198  &
 \\
254 &          &     &     &254     &  254 &   254   &  254  &   254  &
252 \\
268 & 263$\pm$5& 264 & 262 &(268)   &      &         &       &   268  &
 \\
338 & 341$\pm$5& 344 & 342 &338(356)&  338 &   338   &  338  &   338  &
332 \\
440 & 443$\pm$5& 442 & 442 &440     &  440 & 438,440 &  440  &   440  &
428 \\
    &          &     &     &        &      &   542   &  542  &        &
540 \\
556 & 557$\pm$5& 554 & 552 &562     &  556 &   556   &  556  &   562  &
 \\
676 &          & 680 &     &        &  676 &   676   &  676  &        &
670 \\
694 &700$\pm$15&     & 695 &704     &      &         &       &   694  &
 \\
    &          &     &     &        &      &   758   &  748  &        &
 \\
832 &840$\pm$15& 800 & 822 &852     &  832 &   832   &  832  &   832  &
820 \\
912 &          &     & 902 &        &      &   912   &  912  &        &
 \\
1012&1040$\pm$20&970 &1025 &        &      &  1074   & 1074  &  1012  &
990 \\
1100&          &1120 &     &        &      &  1100   & 1100  &  1100  &
 \\
1206&1220$\pm$20&    &     &        &      &         &       &  1216  &
1182 \\
1284&          &     &1297 &        &      &  1284   & 1284  &        &
 \\
1314&          &1310 &     &        &      &         &       &  1314  &
 \\
1410&1430$\pm$20&    &     &        &      &         &       &        &
1398 \\
1502&          &1500 &     &        &      &  1502   &  1502 &  1516  &
 \\
 \hline
\end{tabular}
\end{table}

\newpage
%%%%%%%%%%%%%%%%%%%%%%% Table 6 %%%%%%%%%%%%%%%%%%%%%%%%%%%%%%%%%%%
\setcounter{table}{5}
%%%%%%%%%%%%%%%%%%%%%%%%%%%%%%%%%%%%%%%%%%%%%%%%%%%%%%%%%%%%%%%%%%%%%%
%%%%%%%%%%%%%%%%%%% Table 6   %%%%%%%%%%%%%%%%%%%%%%%%%%%%%%%%%%%%%%%%

\begin{table}

\caption{ Magic numbers provided by the 3-dimensional $q$-deformed harmonic 
oscillator (Table 1), reported in column 1, are compared to the experimental 
data for clusters of Li [10] (column 2), 
Na [6] (column 3), K [12] (column 4), Rb [13] (column 5), Cs
[7,14] (column 6), Cu [16] (column 7), Ag ([18] in column 8, [16] in
column 9), and Au [16] (column 10). See text for discussion. }

\bigskip

\begin{tabular}{c c c c c c c c c c}
\hline
 th.& exp. & exp.    & exp.&exp.&exp.    & exp. & exp.   & exp.    & exp.   \\
present& Li & Na     & K   &Rb&  Cs      &  Cu  &  Ag    &  Ag     &   Au \\
 Tab.1& Ref.[10]& Ref.[6] &Ref.[12]& Ref.[13]& Ref.[7,14]& Ref.[16] & 
Ref.[18] & Ref.[16] & Ref.[16] \\  
\hline
  2  &    &    2      &   2 & 2&    2     &  2   &         &   2     &   2 \\
  8  &    &    8      &   8 & 8&    8     &  8   &     8   &   8     &   8 \\
(18) &    &   18      &     &18&   18     &      &         &         &     \\
 20  &    &   20      &  20 &20&   20     & 20   &    20   &  20     &  20 \\
 34  &    &   34      &     &34&   34     & 34   &    34   &  34     &  34 \\
 40  &    &   40      &  40 &40&   40     & 40   &   (40)  &  40     &     \\
 58  &    &   58      &  58 &  &   58     & 58   &    58   &  58     &  58 \\
 92  &  93&  90,92    &     &  &   92     & 92   &    92   &    92   &  92 \\
138  & 134&  138      &     &  &  138     &138   &    138  &      138& 138 \\
198  & 191&  198$\pm$2 &    & & 198$\pm$2 &      &186$\pm$4&   198   &     \\
254  &    &           &     & &           &      &         &         &     \\
268  & 262&  263$\pm$5 &    & & 263$\pm$5 &      &268$\pm$5&         &     \\
338  & 342& 341$\pm$5 &     & & 341$\pm$5 &      &338$\pm$15&        &     \\
440  & 442& 443$\pm$5 &     & & 443$\pm$5 &      &440$\pm$15&        &     \\
556  & 552& 557$\pm$5 &     & & 557$\pm$5 &      &         &         &     \\
676  &    &           &     &  &          &      &         &         &     \\
694  & 695&  700$\pm$15&    & & 700$\pm$15&      &         &         &     \\
832  & 822&  840$\pm$15&    & & 840$\pm$15&      &         &         &     \\
912  & 902&           &     &  &          &      &         &         &     \\
1012 &1025& 1040$\pm$20&    & &1040$\pm$15&      &         &         &     \\
1100 &    &           &     &  &          &      &         &         &     \\
1206 &    & 1220$\pm$20&    & &           &      &         &         &     \\
1284 &1297&           &     &  &          &      &         &         &     \\
1314 &    &           &     &  &          &      &         &         &     \\
1410 &    &  1430$\pm$20&   & &           &      &         &         &     \\
1502 &    &           &     &  &          &      &         &         &     \\
\hline
\end{tabular}
\end{table}

\newpage
%%%%%%%%%%%%%%%%%%%%%%% Table 7 %%%%%%%%%%%%%%%%%%%%%%%%%%%%%%%%%%
\setcounter{table}{6}

%%%%%%%%%%%%%%%%%%% Table 7 %%%%%%%%%%%%%%%%%%%%%%%%%%%%%%%%%%%%%%%%

\begin{table}

\caption{Magic numbers provided by the 3-dimensional $q$-deformed harmonic 
oscillator of Table 1 with energy gap $\delta =0.26$ (column 1)  and of
Table 2
(column 2),  are compared to the experimental 
data for Zn clusters [19] (column 4) and Cd clusters [19] (column 5), 
as well as to the theoretical predictions of a potential intermediate 
between the simple harmonic oscillator and the square well potential 
[19] (column 3). In addition, the magic numbers provided by the 
3-dimensional $q$-deformed harmonic oscillator of Table 3 (reported in column 
6) are compared to the experimental data for Al [20] (column 7) and In 
[20] (column 8). See text for discussion.  }

\bigskip

\begin{tabular}{c c c c c c c c}
\hline
 th.   & th.   & th.     &exp.    &exp.     &th.     & exp.      &  exp. \\
present&present&         & Zn     & Cd      &present &    Al     &  In    \\
Tab. 1 &Tab. 2 &Ref.[19] &Ref.[19]& Ref.[19]& Tab. 3 & Ref.[20] & Ref.[20] \\
\hline
  2 &  2&     &    &    &    2      &           &        \\
  8 &  8&     &    &    &    8      &           &        \\
 20 & 20&  20 & 20 & 20 &   20      &           &        \\
 34 & 34&  34 &(36)&(36)&   34      &           &        \\
 40 & 40&  40 & 40 & 40 &   40      &           &        \\
 58 & 58&  58 & 56 & 56 &   58      &           &        \\
    &   &     &(60)&(60)&           &           &        \\
    &   &  68 &(64)&(64)&           &           &        \\
 70 & 70&  70 & 70 & 70 &           &           &        \\
    &   &     &(80)&(80)&           &           &        \\
    &   &     &(82)&    &           &           &        \\
 92 & 92&  92 & 92 & 92 &     92    &           &        \\
106 &106& 102 &108 &108 &           &           &        \\
    &112& 112 &(114)&    &           &           &        \\
    &   &    &(120)&(120)&           &           &        \\
138 &138& 138 &138 &138 &  138      & 138       &138     \\
    &   &     &    &    &           & 164       &        \\
    &   &     &    &    &  186      &           &        \\
    &   &     &    &    &           & 198       & 198    \\
    &   &     &    &    &  254      &           &252     \\
    &   &     &    &    &   338     & 336       &        \\
    &   &     &    &    &   398     &           &        \\
    &   &     &    &    &   440     & 438       &        \\
    &   &     &    &    &   486     & 468$\pm$6 &        \\
    &   &     &    &    &   542     & 534$\pm$6 &        \\
    &   &     &    &    &   612     & 594$\pm$6 &        \\
    &   &     &    &    &   676     & 688$\pm$6 &        \\
    &   &     &    &    &   748     & 742$\pm$6 &        \\
    &   &     &    &    &   832     & 832$\pm$10&        \\
    &   &     &    &    &   890     &           &        \\
    &   &     &    &    &   912     & 918$\pm$10&        \\
    &   &     &    &    &  1006     &1000$\pm$10&        \\
    &   &     &    &    &  1074     &           &        \\
    &   &     &    &    &  1100     &1112$\pm$10&        \\
    &   &     &    &    &  1206     &1224$\pm$10&        \\
\hline
\end{tabular}
\end{table}

\end{document}